\documentclass{article}
\usepackage{graphicx} % Required for inserting images
\usepackage[hyphens]{url}
\usepackage{hyperref}
\usepackage[hyphenbreaks]{breakurl}
\usepackage{amsmath}
\usepackage{xcolor}
\usepackage{multirow}
\usepackage{authblk}

\usepackage{pifont}% http://ctan.org/pkg/pifont
\usepackage[symbol]{footmisc}

\usepackage{dcase2022_techrep,url,times,booktabs, tabularx}

\makeatletter
\newcommand\footnoteref[1]{\protected@xdef\@thefnmark{\ref{#1}}\@footnotemark}
\makeatother

\title{DCASE_audiocaptioning_report_plain}
\author{xhajek9}
\date{May 2023}

\renewcommand{\thefootnote}{\fnsymbol{footnote}}

\title{A Whisper transformer for audio captioning trained with synthetic captions and transfer learning}

\name{Marek Kadlčík$^{*,1,2}$,
      Adam Hájek$^{*,1,2}$,
      Jürgen Kieslich$^{2}$, 
      Rados\l{}aw Winiecki$^{2,3}$,
      }
    
\address{$^1$ Faculty of Informatics, Masaryk University, Brno, Czech Republic, \{kadlcik, ahajek\}@mail.muni.cz \\          
        $^2$ Student at Johannes Kepler University, Linz, Austria\\
        $^3$ Student at Politechnika Poznańska, Poznan, Poland\\
 }

\begin{document}

\ninept
\maketitle
{\let\thefootnote\relax\protect\footnotetext{* These authors contributed equally}}

\begin{sloppy}

\begin{abstract}
The field of audio captioning has seen significant advancements in recent years, driven by the availability of large-scale audio datasets and advancements in deep learning techniques. In this technical report, we present our approach to audio captioning, focusing on the use of a pretrained speech-to-text Whisper model and pretraining on synthetic captions. We discuss our training procedures and present our experiments' results, which include model size variations, dataset mixtures, and other hyperparameters. Our findings demonstrate the impact of different training strategies on the performance of the audio captioning model. Our code and trained models are publicly available on GitHub and Hugging Face Hub.
\end{abstract}


\section{Introduction}
Audio captioning is a challenging task that involves generating natural language descriptions for audio content. As opposed to speech-to-text, captioning describes the content of audio clips, such as prominent sounds, music, or environmental noises. This task has numerous practical applications, such as providing access to audio information for people with hearing impairments or improving the searchability of audio content. In recent years, significant progress has been made in audio captioning, thanks to the availability of large-scale audio datasets and advances in deep learning techniques.

In this technical report, we present our approach to audio captioning, which builds on recent developments in the field. The rest of this report is organized as follows. First, we describe the datasets we used for training and validation. Then, we present our model architecture and preprocessing. In the following section, we explain our pretraining and finetuning procedure. Next, we describe several experiments and present the results. Finally, we present the final solution and evaluation scores, summarise our findings, and suggest directions for future research.

As one of the outcomes of our work, we release the code on GitHub\footnote{\url{https://github.com/prompteus/audio-captioning}} and publish trained models on Hugging Face Hub\footnote{\url{https://huggingface.co/MU-NLPC}}.

\section{Data}
We used three sources of data: AudioSet, AudioCaps, and Clotho.

\subsection{Clotho v2.1}

Clotho~\cite{clotho} is a high-quality audio captioning dataset. It contains around 7K audio clips in total, and each clip was manually labeled by five human annotators. Captions are 8-20 words long, and clips are no longer than 30 seconds. For more information about diversity, annotation process, and data quality, see~\cite{clotho}.

\subsection{AudioCaps}

AudioCaps~\cite{audiocaps} is a human-annotated audio captioning dataset. It contains about 46K audio clips in total. The validation and the test split contain five captions for each clip. The train split contains one caption per clip. For more detail, see~\cite{audiocaps}.

\subsection{AudioSet}

AudioSet~\cite{audioset} is a large-scale multi-class and multi-label audio classification dataset containing around 2 million clips. Its audio clips are a superset of AudioCaps. Both datasets need to be scraped from YouTube, and the availability of clips might vary over time.

AudioSet is imbalanced, with around half examples containing the label \textit{Speech} and half containing the label \textit{Music}. We gather the dataset and create a subset that satisfies several conditions:
\begin{enumerate}
    \item Audio clips are available during the scraping
    \item Any leakage between AudioSet and AudioCaps is prevented. If a clip appears in both AudioCaps and our AudioSet subset, it must appear in the same split in both (train, valid, or test).
    \item The label distribution better matches our end use case. Specifically, each class was intended to appear 80 times in the subset if the class is related to music (158 classes, mostly musical instruments and genres) and 500-600 times otherwise (remaining 474 classes). However, note that the data is multi-label, so exact balancing is practically impossible. Clips not containing speech or music were preferred when sampling among multiple choices. 
\end{enumerate}

The created AudioSet subset contains over 130K clip-label pairs and is available with the subset selection algorithm in our repository for better reproducibility. 

In order to train captioning models with AudioSet labels, we create synthetic captions from the labels. Each AudioSet class is associated with a human-written name, and all classes are organized in a hierarchical ontology~\cite{audioset}. To convert a label consisting of classes $C_1, C_2, ..., C_n$, we first prepend each class $C_i$ with its direct parent class in the hierarchy, drop all duplicates, map each class to its name, and finally concatenate them with commas. An example can be seen in \ref{fig:synth-caption}
\
\begin{figure}[htb]
    {\ttfamily{[/m/012n7d, /g/11b630rrvh] $\longrightarrow$ "emergency vehicle, siren, ambulance (siren), domestic sounds - home sounds, whistle, kettle whistle"}}

    \label{fig:synth-caption}
    \caption{Creating a synthetic caption from an AudioSet label}
\end{figure}

\section{Model}

As a base for our model, we decided to use a pretrained Whisper model by OpenAI~\cite{whisper}. Whisper is an encoder-decoder transformer architecture trained on 680K hours of multi-lingual speech-to-text data. The assumption is to utilize the Whisper's ability to extract meaningful patterns in audio sequences and generate coherent text. 

\section{Preprocessing}

In order to best exploit the transfer learning, we mimic the Whisper input-output convention. This convention determines the preprocessing pipeline of both audio files and captions.

\subsection{Audio preprocessing}

First, we down-sample the audio signal to 16 kHz. The original sampling rate of the target Clotho dataset is 44 kHz. We subjectively inspected the effect of down-sampling on audio quality and concluded that it does decrease the clarity of high-frequency sounds but doesn't damage the overall audio. Downsampling to 16K was also used for experimental audio captioning models by Kim et al.~\cite{audiocaps} when proposing the AudioCaps dataset. 

The second audio pre-process step is to convert the audio signal to log-mel spectrograms. We use WhisperFeatureExtractor implemented in the transformers library \footnote{\url{https://huggingface.co/docs/transformers/model_doc/whisper}} with the parameters used by the original Whisper model~\cite{whisper}.

\subsection{Caption preprocessing}

All captions are provided with a constant Whisper prefix (specifying the task of \textit{transcription} and \textit{English} output language) used during speech-to-text training by OpenAI. This ensures the pretrained decoder correctly initiates the generation process. In addition, we prepend the caption with our own prefix, informing the model about the source dataset and captioning task (\textit{caption} vs \textit{keywords}).
An example of synthetic captions can be found in Figure \ref{fig:prefixed-captions}
During inference, the red and blue prefix (as seen in Figure \ref{fig:prefixed-captions}) is forced to the beginning of the sequence. This ensures that the model generates the caption in the desired style.

\begin{figure}[htb]
    % \centering
    {
    \ttfamily{\textcolor{red}{<|startoftranscript|>\allowbreak<|en|>\allowbreak<|transcribe|>\allowbreak<|notimestamps|>}\allowbreak\textcolor{blue}{clotho > caption:} The birds are chirping outdoors throughout the background feedback of persistent wind.\textcolor{red}{<|endoftext|>}}
    
    \vspace{1em}
    
    {\textcolor{red}{<|startoftranscript|>\allowbreak<|en|>\allowbreak<|transcribe|>\allowbreak<|notimestamps|>}\allowbreak\textcolor{blue}{audioset > keywords: }music, music mood, scary music\textcolor{red}{<|endoftext|>}}
    
    \vspace{1em}
    
    {\textcolor{red}{<|startoftranscript|>\allowbreak<|en|>\allowbreak<|transcribe|>\allowbreak<|notimestamps|>}\textcolor{blue}{audiocaps > caption: }An engine accelerating ghastly and then idling\textcolor{red}{<|endoftext|>}}
    }

    \caption{Examples of output sequences. The red parts are tokens used by Whisper architecture, and the blue part denotes our own prefix specifying the source dataset and captioning task.}
    \label{fig:prefixed-captions}
\end{figure}

\section{Training}

\subsection{Audio captioning (AC) pretraining}
In the AC pretraining phase, we aim to introduce various sounds from different domains to our model. We used all 3 datasets: AudioSet with synthetic captions, AudioCaps, and Clotho. The motivation for mixing the datasets during pretraining was (1) to expose the model to a diverse set of audio sounds via AudioSet and (2) to preserve the models' ability to generate fluent sentences via AudioCaps, and (3) to try to speed up the finetuning convergence by exposing the model to Clotho during pretraining. We experimentally select a suitable mixture ratio in Section \ref{sec:experiments}.

We were pretraining on a single Nvidia A100 GPU, and this phase took 34 hours for the largest model (13500 steps with batch size 32).

\subsection{Finetuning}

The finetuning phase focused on learning the specific style of Clotho captions, so we used the Clotho dataset only. The largest model with the final training setting converged in 5 hours of training on a single Nvidia A100 GPU when exposed to Clotho (2200 steps with batch size 32).

\subsection{Augmentations}

During finetuning, we applied the following waveform augmentations:
\begin{enumerate}
    \item adding Gaussian noise
    \item temporal shift (with rollover)
    \item adjusting gain (with clipping values to range -1 to 1)
\end{enumerate}
Each augmentation was applied with a probability of 0.3 independently. For specific implementation details, see our repository.

\section{Experiments}
\label{sec:experiments}

We experimented with different pretraining setups, model sizes, learning rates, dataset mixtures, layer freezing, low-rank adaptation, and decoding strategies.

\subsection{Speech-to-text (STT) pretraining}

In preliminary experiments, using STT-pretrained weights greatly improved the generative capability of the model and convergence speed during training compared to using random initialization. The same holds for the STT-pretrained decoder and randomly initialized encoder setting. In all further experiments, we initialize the models with speech-to-text checkpoints.

\subsection{Audio captioning (AC) pretraining}

We experimented with three pretraining dataset mixtures AudioSet:AudioCaps:Clotho - 1:1:0, 3:1:0, and 12:3:1. 

The ratio 1:1:0 contains the same number of synthetic AudioSet captions and human-written AudioCaps captions in each batch (on average). This setting seemed to overfit AudioCaps and underfit Audioset.

The ratio 3:1:0 includes more synthetic captions in each batch and makes one epoch end on AudioSet and AudioCaps datasets approximately at the same time. This measure slowed overfitting on AudioCaps down and helped the model incorporate more general audio recognition from AudioSet.

We also tried to add a bit of Clotho captions to the pretraining phase in a ratio of 12:3:1. The ratio keeps the proportion of synthetic and human-written captions 3:1 used in the previous mixture and still roughly aligns\footnote{note that Clotho clips are treated as five separate data points during training because of multiple annotations.} with the dataset sizes. In our experience, mixing the Clotho dataset in the pretraining mixture greatly improves finetuning convergence speed.

Table \ref{tab:size-and-pretraining} shows that AC pretraining consistently improves the end task's performance across used model sizes.

\begin{table}[htb]
    \resizebox{\columnwidth}{!}{%
      \begin{tabular}{ |l|r|r| }
      \hline
        Model size    & Finetuning only & AC pretraining + finetuning \\
      \hline
        \textit{whisper-tiny}  & 0.1883 & 0.2101 \\
        \textit{whisper-small} & 0.2186 & 0.2515 \\
        \textit{whisper-large-v2} & 0.2269 & 0.2620 \\
      \hline
      \end{tabular}
    }
    \caption{Comparison of model sizes and the effect of AC pretraining on a mixture of AudioSet and AudioCaps (3:1:0). No augmentations were used. The reported value is a SPIDEr on a custom Clotho validation split\protect\footnotemark{} with greedy decoding.}
    \label{tab:size-and-pretraining}
\end{table}

\addtocounter{footnote}{-1}
\footnotetext[\thefootnote]{\label{note:valid-set}Due to the computational overhead of autoregressive text generation, for this experiment, we shrunk the size of the validation split to 200 and the evaluation split to 400 clips, moving the rest to the train split.}
\addtocounter{footnote}{1}

\subsection{Model size}

We compared 3 model sizes: \textit{whisper-tiny} with 39M parameters, \textit{whisper-small} with 244M parameters, and \textit{whisper-large-v2} with 1550M parameters. 
As Table \ref{tab:size-and-pretraining} shows, model size consistently helps in both finetuning only and AC pretraining + finetuning scenarios.

\subsection{Learning rate}

We tried two learning rates: $4\mathrm{e}{-6}$ and $2\mathrm{e}{-5}$ during finetuning of the pretrained \textit{whisper-small}. Finetuning with a lower learning rate was more stable. Since the training was costly, we assumed that the same would also hold for \textit{whisper-large-v2}. 

\subsection{Finetuning mixture}

In our experiments, finetuning the whole model resulted in overfitting and poor generalization. We attribute that to the large capacity of the models (even \textit{whisper-small}) compared to the modest size of the Clotho dataset.

As a countermeasure, we tried mixing Clotho dataset during finetuning with AudioCaps, or even AudioSet. The intuition is that a robust model should perform well on audio instances outside of the Clotho train split.

We decided to try two ratios. Ratio 1:1:2 means that for each Clotho instance, one out-of-domain instance is added, either from AudioSet or AudioCaps. With a second ratio, 3:1:1, we extend the 3:1:0 pretraining ratio with one piece of Clotho captions. 

However, we found that finetuning mixture hurts the performance on Clotho. Comparison can be found in Table \ref{tab:finetuning-mixture}.

\begin{table}[htb]
    \centering
    \begin{tabular}{ |c|c| }
    \hline
        dataset ratio & SPIDEr \\
        \hline
        0:0:1 & 0.2515 \\
        1:1:2 & 0.2319 \\
        3:1:1 & 0.2338 \\
     \hline
    \end{tabular}
    \caption{Comparison of different dataset mixtures during the finetuning phase. The model a \textit{whisper-small} pretrained with mixture 3:1:0. Augmentations were not used. The reported value is a SPIDEr on a custom Clotho validation \protect\footnotemark{} split with greedy decoding.}
    \label{tab:finetuning-mixture}
\end{table}

\addtocounter{footnote}{-1}
\addtocounter{footnote}{-1}
\footnotetext{See footnote \ref{note:valid-set}}
\addtocounter{footnote}{1}

\subsection{Layer freezing}

Our second approach for preventing memorization of Clotho during finetuning was to freeze parts of the pretrained model, thus decreasing its capacity.

Our preliminary experiments suggested that finetuning around 25\% of parameters in \textit{whisper-small} and \textit{whisper-large} works best. We decided to finetune all \textit{fc1} (23\% parameters in small, 27\% in large) layers to have the updated parameters distributed over the whole architecture.

\subsection{Low-rank adaptation (LoRa)}

Our third approach for preventing overfitting during finetuning was using a low-rank adaption finetuning~\cite{lora}. This parameter-efficient finetuning method freezes a linear layer and extends its functionality with a pair of layers with a small intermediate dimension (rank). Only the small layers are trained, which allows for tuning a great portion of the original model while keeping the number of trainable parameters low. 

There are two critical hyperparameters: the selection of adapted layers and the rank. 
Based on preliminary experiments with a fixed rank of 64, we decided to select all linear layers. Then, we tried to finetune a \textit{whisper-large-v2} pretrained on a 12:3:1 mixture with a LoRA rank of 64 and 256, giving us SPIDEr of 25.18 and 25.12, respectively, on the Clotho validation split, which is comparable but not exceeding a simple model freezing strategy in a same setup with a score of 25.46.

\subsection{Decoding strategy}

An important set of hyperparameters for generative language models are those defining autoregressive decoding strategy. We tried: \textit{greedy decoding}, \textit{multinomial sampling}, \textit{beam-search}, \textit{multinomial beam-search}, \textit{diverse beam-search} and \textit{contrastive search} with each strategy having its own hyperparameters to tune. 

We found that \textit{contrastive search}, a recently proposed promising technique for generating coherent nonrepetitive text using large language models, is not suitable for caption generation. The central hyperparameter $\alpha$, responsible for penalizing the model for repetitions, was inversely correlated with the SPIDEr metric (i.e., the less contrastive a contrastive search was, the better it performed).

Some of the listed strategies rely on token sampling, which introduces non-determinism. While a certain amount of randomness can be useful for general text generation, on such short pieces of text as our captions are, it introduces unwanted variation in performance.

In preliminary experiments, the best performing strategy was \textit{beam-search decoding}. We then tried twelve different values between 3 and 50 as the number of beams on a small validation subset, with values between 5 and 15 giving the highest score. We compare 1, 5, and 10 beams on the whole validation set on all model sizes in Table \ref{tab:num-beams}. We chose the \textit{beam-search} with 5 beams as a preferred decoding strategy.

\begin{table}[htb]
    \centering
    \begin{tabular}{ |l|r|r|r| }
    \hline
         & 1 beam & 5 beams & 10 beams \\
        \hline
        \textit{whisper-tiny}     & 0.2024 & 0.2180 & 0.2146 \\
        \textit{whisper-small}    & 0.2452 & 0.2465 & 0.2429 \\
        \textit{whisper-large-v2} & 0.2546 & 0.2629 & 0.2606 \\
     \hline
    \end{tabular}
    \caption{Comparison of different numbers of beams. The models were pretrained on mixture 12:3:1 and then finetuned on Clotho with all layers except all \textit{fc1} frozen. The reported score is SPIDEr on the Clotho validation split.}
    \label{tab:num-beams}
\end{table}

\section{Evaluation}

For the final evaluation and DCASE2023 challenge submission, we chose three models with the same pretraining and finetuning strategy, differing only in size. The three models are based on pretrained speech-to-text checkpoints of \textit{whisper-tiny}, \textit{whisper-small}, and \textit{whisper-large-v2}. They are then pretrained on a mixture of AudioSet subset with synthetic captions, AudioCaps, and Clotho, with a ratio of 12:3:1. Finally models are finetuned on Clotho only with all layers except all \textit{fc1} frozen. The predictions on the development-evaluation split were generated using beam search with five beams.

The performance of the submitted models on the Clotho development-evaluation dataset can be found in Table \ref{tab:results}.

\begin{table}[htb]
    \resizebox{\columnwidth}{!}{%
        \begin{tabular}{ |l|c|c|c| }
        \hline
             & \textit{whisper-tiny} & \textit{whisper-small} & \textit{whisper-large-v2} \\
            \hline
             SacreBLEU & 13.77 & 15.76 & 16.50 \\
             METEOR & 0.3452 & 0.3781 & 0.3782\\
             CIDEr & 0.3404 & 0.4142 & 0.4331 \\
             SPICE & 0.1077 & 0.1234 & 0.1257 \\
             SPIDEr & 0.2240 & 0.2687 & 0.2794 \\
         \hline
        \end{tabular}
    }
    \caption{The performance of submitted models on Clotho development-evaluation dataset.}
    \label{tab:results}
\end{table}

\section{Discussion and future work}

We identify several areas for potential improvements of our approach: data scaling, architecture, and using stronger supervision.

\subsection{Dataset scaling laws}

In this work, we used datasets with both synthetic and human-written captions. Analyzing how the size of each dataset influences the quality of the model could quantify the value of high-quality human-written captions in comparison to synthetic labels. One could further scale the size of the synthetic captions dataset, for example, by including the FSD50K dataset ~\cite{fsd50k}. Another approach for scaling the pretraining data would be to use a large-scale weak captions dataset, such as WavCaps~\cite{wavcaps}.

\subsection{Architecture}

In audio processing, convolutional neural networks (CNNs) are a popular architecture family. Comparing the effectiveness of a transformer encoder and a CNN encoder could help establish the preferred architecture for the audio captioning task.

\subsection{Stronger supervision}

The available pretrained speech-to-text Whisper checkpoints natively understand timestamps as a part of labels. We believe that the method for creating synthetic captions used in this work can be naturally extended to temporally-strong labels, available, for example, in the AudioSet Strong dataset published by Hershey et al.~\cite{strong-labels}. We suspect that pretraining on temporally-strong synthetic captions could yield additional improvements in audio captioning.

\section{Conclusion}

We trained a transformer encoder-decoder Whisper model for audio captioning and conducted experiments to compare the training regime.

We found that using checkpoint pretrained on speech-to-text helps and further pretraining on a mixture of synthetic captions and human-written from a different source also improves performance compared to finetuning alone. We also observed that larger model sizes result in better performance.

Furthermore, we experiment with balancing the dataset mixture during the pretraining. We found that a higher emphasis on a large dataset with synthetic captions helps mitigate overfitting. During finetuning, we tried several regularization strategies to prevent overfitting, with simple layer freezing performing best. 

Overall, our approach demonstrates promising results in generating natural language descriptions for audio content. However, there is still room for further exploration and improvement. Our findings contribute to the advancement of audio captioning and provide insights for researchers and practitioners working in the field.

\section{Acknowledgement}

Computational resources were supplied by the Natural Language Processing Centre at the Faculty of Informatics at Masaryk University. 

This research was conducted as part of the course \textit{Special Topics: Machine Learning and Audio: A challenge} at Johannes Kepler University we attended as students.

\bibliographystyle{IEEEtran}
\bibliography{bibliography}

\end{sloppy}
\end{document}